\documentclass[letterpaper]{article} 
\usepackage{aaai2026} 
\usepackage{times} 
\usepackage{helvet}  
\usepackage{courier}  
\usepackage[hyphens]{url}  
\usepackage{graphicx} 
\usepackage{booktabs} 
\usepackage{natbib}  
\usepackage{algorithm}
\usepackage{algorithmic}
\usepackage{newfloat}
\usepackage{listings}
\usepackage{subcaption}
\usepackage{xcolor} 
\usepackage{enumitem}
\usepackage{amsmath}
\usepackage{amssymb}
\usepackage{caption}
\usepackage{amsfonts}
\usepackage{multirow} 
\usepackage{placeins} 
\usepackage{array}
\usepackage{float}

\DeclareCaptionStyle{ruled}{labelfont=normalfont,labelsep=colon,strut=off} 
\floatstyle{ruled}
\newfloat{listing}{tb}{lst}{}
\floatname{listing}{Listing}
\title{RABot: Reinforcement-Guided Graph Augmentation for \\ Imbalanced and Noisy Social Bot Detection}
\author {
    Longlong Zhang\textsuperscript{\rm1,\rm2},
    Xi Wang\textsuperscript{\rm3},
    Haotong Du\textsuperscript{\rm4},
    Yangyi Xu\textsuperscript{\rm1},
    Zhuo Liu\textsuperscript{\rm1},
    Yang Liu\textsuperscript{\rm1,\rm2}\thanks{Corresponding author.}
}
\affiliations {
    \textsuperscript{\rm 1}School of Artificial Intelligence, Optics and Electronics (iOPEN), Northwestern Polytechnical University\\
    \textsuperscript{\rm 2}Shenzhen Research Institute, Northwestern Polytechnical University\\
    \textsuperscript{\rm 3}School of Automation Science and Engineering, Xi’an Jiaotong University\\
    \textsuperscript{\rm 4}School of Computer Science, Northwestern Polytechnical University\\
    \{zhanglonglong, duhaotong, xu\_yangyi, zhuoliu02\}@mail.nwpu.edu.cn, vancywangxi@ust.hk, yangliuyh@gmail.com
}

\begin{document}

\maketitle

\begin{abstract}

Social bot detection is pivotal for safeguarding the integrity of online information ecosystems. Although recent graph neural network (GNN) solutions achieve strong results, they remain hindered by two practical challenges: (i) severe class imbalance arising from the high cost of generating bots, and (ii) topological noise introduced by bots that skillfully mimic human behavior and forge deceptive links. We propose the \textbf{R}einforcement-guided graph \textbf{A}ugmentation social \textbf{Bot} detector (RABot), a multi-granularity graph-augmentation framework that addresses both issues in a unified manner. RABot employs a neighborhood-aware oversampling strategy that linearly interpolates minority-class embeddings within local subgraphs, thereby stabilizing the decision boundary under low-resource regimes. Concurrently, a reinforcement-learning-driven edge-filtering module combines similarity-based edge features with adaptive threshold optimization to excise spurious interactions during message passing, yielding a cleaner topology. Extensive experiments on three real-world benchmarks and four GNN backbones demonstrate that RABot consistently surpasses state-of-the-art baselines. In addition, since its augmentation and filtering modules are orthogonal to the underlying architecture, RABot can be seamlessly integrated into existing GNN pipelines to boost performance with minimal overhead.

\end{abstract}

\section{Introduction}

With social media now serving as the primary platform for public discourse, the infiltration of social bots has become a growing academic and practical concern \cite{Cresci2020ADecade}. Social bots are software agents that imitate human users and can participate in large-scale interactions across online platforms \cite{Ferrara2016Therise}. Empirical studies have linked them to a range of harmful behaviors, including dissemination of disinformation \cite{Shao18hoaxybots}, interference in elections \cite{Ferrara2017Disinformation}, manipulation of public opinion \cite{Cheng2020Dynamic}, and the spread of extremist ideologies \cite{fernandez2021artificial, liu2025efficientedge}. Effective bot-detection techniques are therefore vital for reducing these threats and maintaining trust in online information ecosystems \cite{orabi2020detection}. 

Current social bot detection schemes mainly fall into two categories: feature-based approaches and graph-based approaches~\cite{Alothali2018Detecting}. Feature-based methods examine either user metadata \cite{yang2020scalable} or textual content \cite{hayawi2022deeprobot}. Metadata encompasses numerical attributes and categorical attributes; classifiers differentiate bots from legitimate accounts by contrasting these values \cite{patil2022comprehensive}. Text-oriented models employ word-vector encoders or pretrained language models to process tweets and profile descriptions, thereby extracting latent semantic cues indicative of automated behavior \cite{wu2023bottrinet}. Although informative, such shallow features often fail to unveil \emph{camouflaged bots} \cite{cresci2017paradigm}. Adversaries can replicate large portions of genuine user information and inject only minimal malicious content, effectively narrowing the observable gap between bots and real accounts \cite{yan2021asymmetrical}.

\begin{figure}[t]
    \centering
    \subfloat{%
    \raisebox{0.3em}{
        \includegraphics[width=0.4\linewidth]{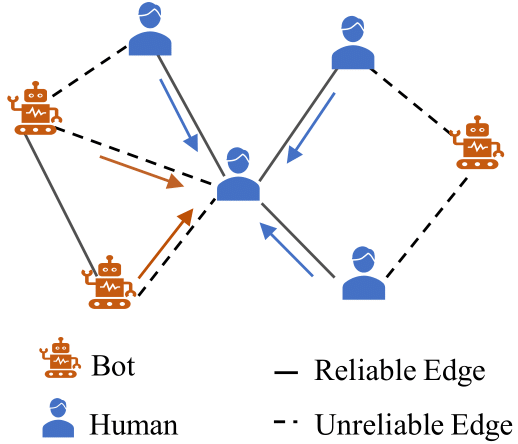}%
    }}
    \hfill            
    \subfloat{%
        \includegraphics[width=0.55\linewidth]{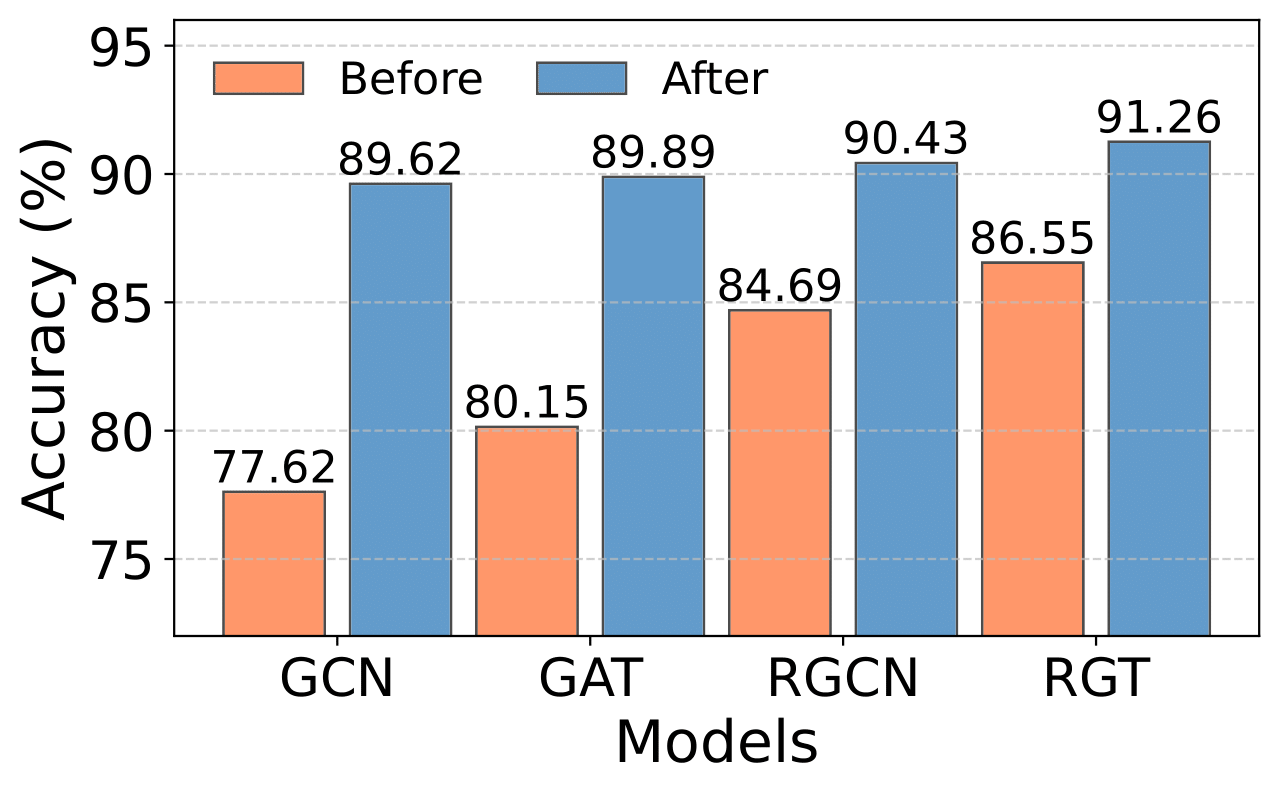}%
    }
    \caption{Unreliable aggregation in social networks (left). Performance variation after removing all unreliable edges on Twibot-20 (right).}
    \label{movitate}
    \vspace{-15pt}
\end{figure}

To address these limitations, recent work has shifted toward graph-based methods that leverage advances in graph neural networks (GNNs)~\cite{feng2022heterogeneity, liu2023botmoe, wu2025search}. These models build on the empirical observation that genuine users form cohesive communities, whereas bots connect in a more random fashion, creating distinctive topological signatures~\cite{yang2013empirical}. 
By explicitly modelling the underlying social-network structure, graph-based detectors can generally surpass feature-based ones \citep{li2024botcl}. Nevertheless, two key challenges persist. \textbf{(i) Class imbalance.} Social graphs contain far fewer bot nodes than human nodes. This skew biases graph-based models toward the majority class and leads to systematic misclassification of bots \citep{liu2022dagad}. Mitigating this extreme imbalance is therefore essential for reliable bot detection. \textbf{(ii) Noisy or spurious edges.} Social graphs often include numerous unreliable interactions, for example, as illustrated in Figure~\ref{movitate}, edges linking a human account to a bot~\cite{Cresci2020ADecade, feng2021twibot}. During message aggregation, these edges introduce bot-generated features into otherwise benign neighborhoods; deeper GNN layers propagate and amplify this noise, ultimately degrading the detection performance.

To tackle the foregoing challenges, we present the \textbf{R}einforcement-guided graph \textbf{A}ugmentation social \textbf{Bot} detector (RABot), a graph-enhanced detection framework that addresses class imbalance and topological noise in a unified manner. RABot first augments the minority class through a neighborhood-aware oversampling module that interpolates new samples along local feature manifolds. It then refines the social graph by estimating edge reliability with a reinforcement-learning agent and pruning low-confidence ties, allowing the subsequent GNN to propagate information over trustworthy neighborhood only. In such a manner, RABot can effectively mitigate the impact of class imbalance and topological noise, thereby significantly improving detection accuracy and robustness. Our main contributions are as follows:

\begin{itemize}[leftmargin=*]
    \item \emph{Neighborhood-aware oversampling.}  
          We introduce a dynamic oversampling scheme that exploits local feature distributions to synthesize minority-class nodes, thus preserving decision boundaries under severe class imbalance.
    \item \emph{Reinforcement-guided edge filtering.}  
          We propose an adaptive edge-pruning mechanism that combines similarity metrics with reinforcement learning to detect and remove spurious or camouflaged connections during GNN aggregation, substantially reducing topological noise.
    \item \emph{Comprehensive evaluation.}  
          RABot is validated on three widely used social bot datasets, where it consistently surpasses the state-of-the-art detectors. Ablation studies also confirm the individual contributions of the oversampling and edge-filtering modules.
\end{itemize}

\section{Related Work}

\subsection{Social Bot Detection} 

Social bot detection aims to identify automated bot accounts on social networking platforms. Early social bot detection models primarily depended on manually designed and extracted features, leveraging user metadata \cite{alothali2021hybrid} and tweets \cite{mazza2019rtbust} in conjunction with traditional classifiers for bot detection. With the rapid advancement of deep learning, neural network-based methods have increasingly been applied to social bot detection tasks, such as LSTM \cite{ alkahtani2021botnet}, GAN \cite{stanton2019gans} and various pre-trained language models \cite{Kumar2021Content} to capture semantic inconsistencies in user information. However, with the advancement of bot anthropomorphism technology, methods that rely solely on user data or tweet information are gradually losing effectiveness. To overcome this limitation, researchers have proposed various methods that integrate user profile information with tweet data \cite{feng2021satar} to enhance bot detection. With the development of GNNs, researchers have gradually incorporated the topological structure of social networks into bot detection tasks \cite{cai2024lmbot, wei2025cagcl}. Since then, researchers have focused on the heterogeneity of social networks \cite{feng2022heterogeneity, liu2025diffusionsource}, leading to significant improvements in detection performance. Subsequent studies introduce multi-view contrastive learning, enhancing detection by maximizing inter-view mutual information and incorporating multi-task objectives \cite{yang2024sebot}. Recent studies achieve state-of-the-art results by dynamically modeling social networks and capturing the evolution of user behavior \cite{He2024Dynamicity}. However, existing methods often overlook class imbalance and unreliable connections in social networks, which can lead to biased learning and degraded detection performance.

\subsection{Graph Augmentation} 

Graph augmentation generally refers to optimizing the quality of graph data by modifying the distribution of node features or reconstructing graph structure. Early graph augmentation methods enhance the generalization and robustness of neural networks by optimizing the feature distribution of graph nodes \cite{velickovic2019deep}. For example, some studies \cite{park2021graphens} construct an embedding space to encode node similarities, ensuring the authenticity of synthesized samples. Subsequently, graph augmentation methods focusing on optimizing graph structures \cite{zhao2022learning, Liu2024Diffusion} have been proposed. Among them, contrastive learning methods have become an important paradigm: through topological perturbations \cite{hou2022graphmae}, subgraph comparisons \cite{moren2022graphcl} and heterogeneous relationship mining \cite{yu2022heterogeneous}, structural robustness enhancement is achieved under self-supervised conditions. In recent years, dynamic adaptation-based graph augmentation methods, such as learnable edge weight \cite{shen2023neighbor}, implicit relationship reasoning \cite{wei2024llmrec}, and multi-scale fusion \cite{duan2023graph}, have gained increasing attention. These methods enhance model adaptability to noisy and sparse data through task-oriented structural optimization.

\begin{figure*}[h]
    \centering
    \includegraphics[width=0.955\textwidth]{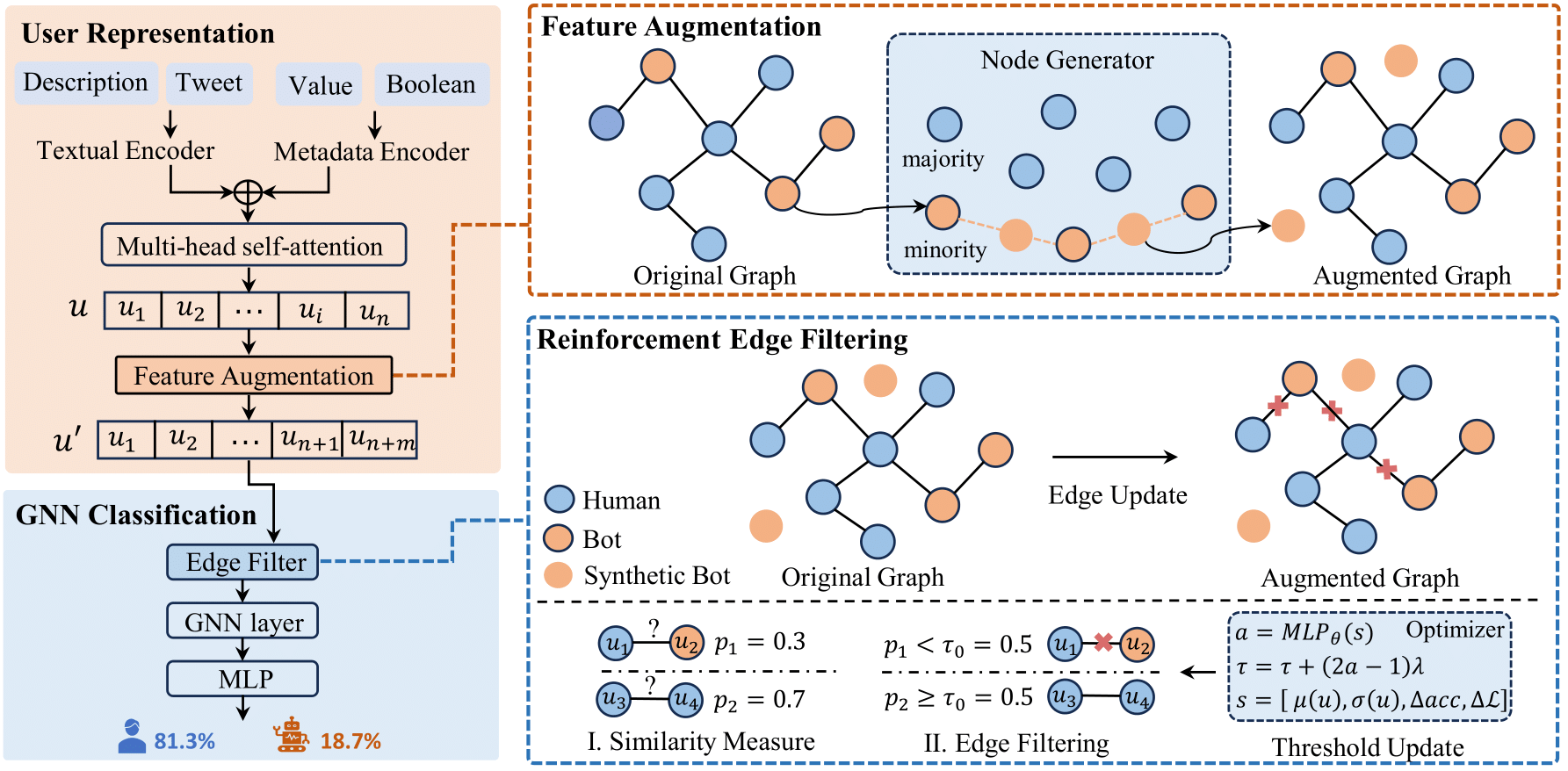}
    \caption{Overall structure of RABot model, which consists of four modules: user information representation module, feature augmentation module, reinforcement edge filtering module and GNN classification module.}
    \label{overall}
\end{figure*}

\section{Problem Definition}

A social network can be represented as a graph \(G=(V,E)\), where \(V=\{v_1,\dots,v_n\}\) denotes the set of users (nodes) and \(E\subseteq V\times V\) denotes the set of social interactions (edges), with $n=|V|$ the number of nodes in $G$. Each node \(v_i\) is associated with a $\kappa$-dimension feature vector \(\mathbf{u}_i\in\mathbb{R}^{\kappa}\) that encodes metadata, textual, and topological cues. \textbf{Social bot detection} is formulated as a node-level binary classification task. We seek an encoder
\(f:\mathbb{R}^{\kappa}\!\rightarrow\![0,1]\) such that, for every
user \(v_i\), the predicted score \(\hat{y}_i=f(\mathbf{u}_i)\) is close to the ground-truth label \(y_i\in\{0,1\}\).
The objective is to maximize classification accuracy over all nodes in
\(V\):
\begin{equation}
\max_{f}\;\frac{1}{|V|}\sum_{i=1}^{|V|}\mathbf{1}(\tilde{y}_i = y_i),
\end{equation}
where \(\mathbf{1}(\cdot)\) is the indicator function and $\tilde{y}_i=\mathbf{1}_{\hat{y}_i\geq 0.5}$.

\section{Methodology}
\label{mymethod}

The overall architecture of RABot is illustrated in Figure~\ref{overall}. Our approach adopts a multi-granularity graph enhancement framework that improves representation learning on social networks through a dual collaborative mechanism. First, a feature-augmentation module based on neighborhood propagation synthesizes compensatory samples for the minority class within local subgraph spaces. Next, an edge filtering module that combines similarity metrics with reinforcement learning dynamically prunes noisy connections and reconstructs the graph topology during message passing, guided by edge weight confidence scores. Finally, the refined multimodal features and purified graph structure are fed into a GNN classifier, yielding accurate bot detection.

\subsection{User Information Representation}

User information in social networks can be divided into profile attributes and textual content. The former contains numerical and Boolean fields, whereas the latter comprises the free-form profile description and the sequence of tweets published by the user. We extract four corresponding feature vectors with a multilayer perceptron (MLP) and a pretrained language model (LM)~\cite{liu2019roberta} as:
\begin{equation}
\begin{aligned}
\label{eq1}
\mathbf{u}_i^{v} &= \mathbf{v}_i \mathbf{W}_v ,\quad
\mathbf{u}_i^{b} = \mathbf{b}_i \mathbf{W}_b ,\quad
\mathbf{u}_i^{d} = \operatorname{LM}(d_i) \mathbf{W}_d ,\\
\mathbf{u}_i^{t} &= \bigl(\operatorname{AvePool}\{\operatorname{LM}(t_i^{1}),\ldots,\operatorname{LM}(t_i^{m})\}\bigr) \mathbf{W}_t,
\end{aligned}
\end{equation}
where $\mathbf{u}_i^{v}$ and $\mathbf{u}_i^{b}$ encode the numerical and Boolean attributes of $i$-th user, $\mathbf{u}_i^{d}$ encodes the profile description $d_i$, and $\mathbf{u}_i^{t}$ encodes the $m$ tweets $\{ t_i^1, \dots, t_i^m\}$. The matrices $\mathbf{W}_v$, $\mathbf{W}_b$, $\mathbf{W}_d$, and $\mathbf{W}_t$ are learnable parameters.

To fuse the heterogeneous features while alleviating semantic mismatch among modalities, we apply a multi-head self attention mechanism~\cite{vaswani2017attention}:
\begin{equation}
\begin{aligned}
\label{eq2}
\mathbf{U} &= \{\mathbf{u}_1,\ldots,\mathbf{u}_n \mid \mathbf{u}_i = \operatorname{Concat}(\mathbf{u}_i^{v},\mathbf{u}_i^{b},\mathbf{u}_i^{d},\mathbf{u}_i^{t})\mathbf{W}\} ,\\
\mathbf{Q}_{c} &= \mathbf{U}\mathbf{W}_{c,q},\quad
\mathbf{K}_{c} = \mathbf{U}\mathbf{W}_{c,k},\quad
\mathbf{V}_{c} = \mathbf{U}\mathbf{W}_{c,v} ,\\
\hat{\mathbf{U}} &= \mathop{\Vert}_{c=1}^{C} \operatorname{Softmax}\!\left(\frac{\mathbf{Q}_{c}\mathbf{K}_{c}^{\top}}{\sqrt{d_k}}\right) \mathbf{V}_{c} ,
\end{aligned}
\end{equation}
where $\mathbf{U}$ is the initial user-level embedding, $d_k = \kappa / C$ is the dimension of each head, and $(\mathbf{Q}_{c},\mathbf{K}_{c},\mathbf{V}_{c})$ are the query, key, and value matrices of the $c$-th head obtained through the trainable weights $\mathbf{W}_{c,q}$, $\mathbf{W}_{c,k}$, and $\mathbf{W}_{c,v}$, respectively. The operator $\Vert$ concatenates the outputs of the $C$ heads, yielding the final representation~$\hat{\mathbf{U}}$.

\subsection{Feature Augmentation}

Conventional GNN oversampling methods synthesize nodes in the original feature space~\cite{duan2022anonymity}. Since these artificial nodes have no natural edges, a post-hoc linking step is required, which often introduces noise. Motivated by the approach~\cite{zhao2021graphsmote}, we instead generate minority-class embeddings directly in the latent space, simultaneously balancing the class distribution and avoiding noisy edge construction.

Let $V = (V_1, V_2)$ denote the user set, where $V_1$ and $V_2$ are minority and majority instances, respectively.  
For minority node $v_i \in V_1$, we identify its $k$ nearest neighbors as:
\begin{equation}
\label{eq3}
\Gamma(v_i)=\operatorname{Top-k} \ \!\bigl(\{v_j \in V_1 \mid v_j \neq v_i\}, \quad L(\mathbf{u}_i,\mathbf{u}_j)\bigr),
\end{equation}
where $L(\mathbf{u}_i,\mathbf{u}_j)=\lVert\mathbf{u}_i-\mathbf{u}_j\rVert_2$ is the Euclidean distance between the latent features of $v_i$ and $v_j$.

We then select a random neighbor $v_x \in \Gamma(v_i)$ and create a synthetic node $v_k$ by linear interpolation:
\begin{equation}
\label{eq4}
\mathbf{u}_k = (1-\delta)\,\mathbf{u}_i + \delta\,\mathbf{u}_x,\quad \delta \sim \mathcal{U}(0,1),
\end{equation}
where $\mathbf{u}_k$ is the feature vector of the generated node $v_k$. The generated node $v_k$ inherits the minority label. Such sampling process is repeated until the class sizes are balanced, yielding a set of labelled virtual nodes that equalizes the training distribution.

\subsection{Reinforcement Edge Filtering}

This module is motivated by two practical needs: (i) to suppress edges that obscure the true class structure of the graph, and (ii) to adjust the filtering strength as training proceeds.  
The workflow therefore proceeds in three stages: computing similarity scores, filtering edges under a learnable threshold, and updating that threshold through reinforcement learning.

\paragraph{Node-level similarity.}
A pair of nodes with similar features is likely to share the same class label and to form a reliable edge, whereas a large feature gap suggests the opposite. However, advanced camouflage enables many bots to imitate human behavior, so naive embedding distances no longer suffice. To obtain a more discriminative measure, we pass each node through a two-layer MLP label predictor and compute the $\ell_{1}$ distance between central node $v_i$ and its corresponding neighbor $v_j$ as:
\begin{equation}
\label{eq5}
D^{\,r,(l)}_{i,j}
  =\left\lVert
        \sigma\!\bigl(\operatorname{MLP}^{(l)}(h_{i}^{\,r,(l-1)})\bigr)
     \! - \! \sigma\!\bigl(\operatorname{MLP}^{(l)}(h_{j}^{\,r,(l-1)})\bigr)
    \right\rVert_{1},
\end{equation}
followed by a similarity probability as:
\begin{equation}\label{eq6}
p^{\,r,(l)}_{i,j}=1-D^{\,r,(l)}_{i,j},
\end{equation}
where $\sigma(\cdot)$ is the {Sigmoid} function and the superscript $(l-1)$ denotes features from the previous GNN layer.

\paragraph{Edge filtering.}
To control filtering strength we introduce a trainable threshold $\tau\in(0,1)$, initialized at $0.5$.  
For each edge $(v_i^{\,r},v_j^{\,r})\in\mathcal{E}^{r}$,

\begin{equation}
\label{eq7}
  m^{\,r,(l)}_{i,j} \;=\;
  \begin{cases}
    1,& p^{\,r,(l)}_{i,j}\ge \tau,\\[2pt]
    0,& p^{\,r,(l)}_{i,j}<  \tau,
  \end{cases}
\end{equation}
where $m^{\,r,(l)}_{i,j}=1$ retains the edge and $0$ removes it.  
We supervise this binary decision with a Binary Cross-Entropy (BCE) loss that rewards correct retention of homogeneous edges and removal of heterogeneous ones obeying

\begin{equation}
\label{eq8}
\mathcal{L}_{\text{edge}}^{(l)}
  = \mathbb{E}\!\Bigl[
      \sum_{r=1}^{R}\!
        \bigl(
          \!\!\sum_{\,i,j:\,y_i=y_j} {\!} {\!} {\!} {\!} {\!}-\log p^{\,r,(l)}_{i,j}
         +\!\!\sum_{\,i,j:\,y_i\neq y_j} {\!} {\!} {\!} {\!} {\!}-\log(1-p^{\,r,(l)}_{i,j})
        \bigr)
    \Bigr].
\end{equation}

\paragraph{Adaptive threshold.}
The optimal value of $\tau$ can differ across datasets and even across training epochs.  
To capture this dynamic we update $\tau$ every $T$ epochs with a lightweight reinforcement signal.

\begin{itemize}[leftmargin=*]
\item \textbf{State.}  
      $\mathbf{s}=[\mu(\mathbf{u}),\sigma(\mathbf{u}),\Delta\text{acc},\Delta\mathcal{L}]\in\mathbb{R}^{4}$,  
      where $\mu$ and $\sigma$ are the mean and standard deviation of node features, and $\Delta\text{acc}$ and $\Delta\mathcal{L}$ measure recent changes in accuracy and loss, respectively.
\item \textbf{Policy.}  
      A small MLP with parameters $\theta$ maps the state $\mathbf{s}$ to an action  
      $a=\sigma\!\bigl(\text{MLP}_{\theta}(\mathbf{s})\bigr)\in(0,1)$.
\item \textbf{Action.}  
    Based on the current action $a$, we apply an update on $\tau$ obeying
     \begin{equation}
        \label{eq13}
        \tau \leftarrow \tau+(2 a-1) \lambda \ ,
    \end{equation}
     where $\lambda$ is a step size and $(2a-1)\in(-1,1)$ determines both direction and magnitude.
\end{itemize}

\subsection{GNN Classification}

Having obtained an augmented feature set and a denoised topology, we refine node embeddings with a generic GNN layer. The framework is architecture-agnostic, so any message-passing variant (e.g.\ GCN) can be substituted.  
For relation $r$ at layer $l{+}1$, the aggregation and update steps are:
\begin{equation}
\begin{aligned}
\label{eq9}
\mathbf{a}_{i}^{\,r,(l+1)} &=
    \operatorname{Aggregate}^{(l+1)}
      \bigl(\{\mathbf{v}_{j}^{\,r,(l)} \mid j\!\in\!\mathcal{N}_{r}(i)\},
            \mathbf{v}_{i}^{\,r,(l)}\bigr),\\[4pt]
\mathbf{v}_{i}^{\,(l+1)} &=
    \Bigl.\operatorname{Update}^{(l+1)}
      \bigl(\mathbf{v}_{i}^{\,r,(l)},\mathbf{a}_{i}^{\,r,(l+1)}\bigr)
    \Bigr|_{r=1}^{R},
\end{aligned}
\end{equation}
where $\mathcal{N}_{r}(i)$ is the neighbor set of $v_i$ under relation~$r$,  
$\mathbf{a}_{i}^{\,r,(l+1)}$ is the aggregated message, and  
$\mathbf{v}_{i}^{\,(l+1)}$ is the updated representation.

A linear output layer followed by {Softmax} produces the class probabilities:
\begin{equation}
\label{eq10}
\hat{y}_{i}= \operatorname{Softmax} \!\bigl(\mathbf{W}_{v}\mathbf{v}_{i}+\mathbf{b}_{v}\bigr),
\end{equation}
where $\mathbf{W}_{v}$ and $\mathbf{b}_{v}$ are learnable parameters and $\hat{y}_{i}$ represents the user category predicted by the model. The model is optimized with binary cross-entropy:
\begin{equation}
\label{eq11}
\mathcal{L}_{\mathrm{gnn}}
  = -\sum_{i=1}^{n}\Bigl[
      y_{i}\log\hat{y}_{i}
    +(1-y_{i})\log(1-\hat{y}_{i})
    \Bigr],
\end{equation}
where $y_{i}\in\{0,1\}$ is the ground-truth label.

\subsection{Learning and Optimization}

To stabilize training and balance competing objectives, we adopt a multi-objective optimization scheme that jointly minimizes classification and edge-quality losses.  
The overall loss is defined as:
\begin{equation}
\label{eq12}
\mathcal{L}
   = \lambda_s\,\mathcal{L}_{\text{gnn}}^{\prime}
   + (1-\lambda_s)\,\mathcal{L}_{\text{gnn}}
   + \lambda_e\,\mathcal{L}_{\text{edge}},
\end{equation}
where $\mathcal{L}_{\text{gnn}}^{\prime}$ and $\mathcal{L}_{\text{gnn}}$ are the cross-entropy losses computed on the augmented nodes and the original nodes, respectively.  
The former mitigates class imbalance, whereas the latter preserves baseline discrimination. The edge component $\mathcal{L}_{\text {edge }}=(1 / L) {\small\sideset{}{_{l=1}^{L}}\sum} \mathcal{L}_{\text {edge }}^{(l)}$ is the average layer-wise BCE loss that guides edge filtering, with $L$ denoting the number of GNN layers. Hyperparameters $\lambda_s\!\in\![0,1]$ and $\lambda_e>0$ control the contribution of each term.

\begin{table*}[t]
\small
\renewcommand{\arraystretch}{0.9} 
\centering
\setlength{\tabcolsep}{9pt}  
\begin{tabular}{@{}ll|ll|ll|ll@{}}
\specialrule{0.8pt}{0pt}{0pt}  
\toprule
\multirow{2}{*}{Methods} &
  Dataset &
  \multicolumn{2}{c|}{Cresci-15} &
  \multicolumn{2}{c|}{Twibot-20} &
  \multicolumn{2}{c}{MGTAB} \\ \cmidrule(l){2-8} 
 &
  Metrics &
  \multicolumn{1}{c}{Accuracy} &
  \multicolumn{1}{c|}{F1-score} &
  \multicolumn{1}{c}{Accuracy} &
  \multicolumn{1}{c|}{F1-score} &
  \multicolumn{1}{c}{Accuracy} &
  \multicolumn{1}{c}{F1-score} \\ \midrule

\multirow{4}{*}{Feature-based} &

  Wei \textit{et al.} &
  96.10$\pm$0.75 & 81.68$\pm$0.68 &
  70.23$\pm$0.17 & 53.37$\pm$0.16 &
  \multicolumn{1}{c}{-} & \multicolumn{1}{c}{-} \\

  &
  Varol \textit{et al.} &
  93.25$\pm$0.24 & 94.82$\pm$0.10 &
  78.73$\pm$0.37 & 81.26$\pm$0.42 &
  \multicolumn{1}{c}{-} & \multicolumn{1}{c}{-} \\

  &
  SATAR &
  93.14$\pm$0.51 & 94.87$\pm$0.32 &
  84.50$\pm$0.67 & 86.54$\pm$0.68 &
  \multicolumn{1}{c}{-} & \multicolumn{1}{c}{-} \\

  &
  SGBot &
  77.40$\pm$0.82 & 78.15$\pm$0.13 &
  81.63$\pm$0.55 & 84.91$\pm$0.34 &
 \multicolumn{1}{c}{-} & \multicolumn{1}{c}{-} \\ \midrule

\multirow{8}{*}{Graph-based} &
  GCN &
  96.49$\pm$0.37 & 96.32$\pm$0.20 &
  77.62$\pm$0.85 & 80.36$\pm$0.49 &
  83.61$\pm$1.47 & 78.16$\pm$1.65 \\

  &
  GAT &
  96.34$\pm$0.38 & 96.05$\pm$0.39 &
  80.15$\pm$1.19 & 80.79$\pm$1.18 &
  86.68$\pm$1.31 & 82.68$\pm$1.77 \\

  &
  RGT &
  97.13$\pm$0.33 & 97.74$\pm$0.32 &
  86.55$\pm$0.23 & 87.67$\pm$0.31 &
  89.58$\pm$0.80 & 86.43$\pm$1.12 \\
  
  &
  BotRGCN &
  96.58$\pm$0.68 & 97.34$\pm$0.50 &
  84.69$\pm$0.40 & 85.47$\pm$0.43 &
  89.27$\pm$1.06 & 86.07$\pm$1.49 \\

  &
  BECE &
  \underline{98.73$\pm$0.62} & 98.57$\pm$0.54 &
  87.24$\pm$0.59 & 88.01$\pm$0.73 &
  90.31$\pm$0.64 & 88.10$\pm$0.52 \\

  &
  LMBot &
  98.69$\pm$0.31 & 98.41$\pm$0.56 &
  86.63$\pm$0.43 & 87.25$\pm$0.59 &
  88.78$\pm$0.49 & 86.12$\pm$0.66  \\

  &
  BotDGT &
  98.62$\pm$0.45 & 98.32$\pm$0.19 &
  87.12$\pm$0.41 & 87.82$\pm$0.54 &
  \multicolumn{1}{c}{-} & \multicolumn{1}{c}{-} \\

  &
  SEBot &
  98.67$\pm$0.44 & 98.54$\pm$0.61 &
  87.26$\pm$0.36 & \underline{88.06$\pm$0.45} &
  90.28$\pm$0.47 & 84.98$\pm$0.36 \\ 
  
  \midrule

\multirow{8}{*}{Ours} &
  RABot\,(GCN) &
  98.18$\pm$0.14 & 98.07$\pm$0.28 &
  82.08$\pm$0.17 & 81.81$\pm$0.23 &
  86.42$\pm$0.05 & 82.89$\pm$0.21 \\

  &
  GCN (+) &
  \multicolumn{1}{c}{+1.69} & \multicolumn{1}{c|}{+1.75} &
  \multicolumn{1}{c}{+4.46} & \multicolumn{1}{c|}{+1.45} &
  \multicolumn{1}{c}{+2.81} & \multicolumn{1}{c}{+4.73} \\

  &
  RABot\,(GAT) &
  98.46$\pm$0.22 & 98.28$\pm$0.26 &
  83.98$\pm$0.17 & 83.41$\pm$0.21 &
  89.94$\pm$0.49 & 86.61$\pm$0.43 \\

  &
  GAT (+) &
  \multicolumn{1}{c}{+2.12} & \multicolumn{1}{c|}{+2.23} &
  \multicolumn{1}{c}{+2.83} & \multicolumn{1}{c|}{+1.62} &
  \multicolumn{1}{c}{+3.26} & \multicolumn{1}{c}{+3.93} \\

  &
  RABot\,(RGT) &
  \textbf{99.14$\pm$0.21} & \textbf{98.94$\pm$0.34} &
  \textbf{87.92$\pm$0.12} & \textbf{88.40$\pm$0.28} &
  \underline{90.74$\pm$0.15} & \underline{88.76$\pm$0.16} \\

  &
  RGT (+) &
  \multicolumn{1}{c}{+2.01} & \multicolumn{1}{c|}{+1.20} &
  \multicolumn{1}{c}{+1.37} & \multicolumn{1}{c|}{+0.73} &
  \multicolumn{1}{c}{+1.16} & \multicolumn{1}{c}{+2.33} \\

  &
  RABot\,(RGCN) &
  98.76$\pm$0.42 & \underline{98.72$\pm$0.47} &
  \underline{87.36$\pm$0.28} & 87.12$\pm$0.19 &
  \textbf{91.16$\pm$0.31} & \textbf{89.03$\pm$0.28} \\ 

  &
  BotRGCN (+) &
  \multicolumn{1}{c}{+2.18} & \multicolumn{1}{c|}{+1.38} &
  \multicolumn{1}{c}{+2.67} & \multicolumn{1}{c|}{+1.65} &
  \multicolumn{1}{c}{+1.89} & \multicolumn{1}{c}{+2.96} \\

  \bottomrule
\specialrule{0.8pt}{0pt}{0pt}  
\end{tabular}%
\caption{Performance comparison of varied detection methods on the Cresci-15, Twibot-20, and MGTAB datasets. Each method is executed five times, and we report the mean and standard deviation (mean\,$\pm$\,std\%). ``GNN~(+)\,'' denotes the absolute improvement of RABot\,(GNN) over its backbone GNN, whereas ``--'' indicates that the method cannot be applied to MGTAB. The best and second-best scores are highlighted with \textbf{bold} and \underline{underline}, respectively.}
\label{tab:modelCompare}
\end{table*}

\section{Experiments}

\label{myresult}

\subsection{Experimental Setup}

\textbf{Datasets and Evaluation.} We benchmark our approach on three widely used social bot detection corpora: {Cresci-15} \cite{cresci2015fame}, {Twibot-20} \cite{feng2021twibot}, and {MGTAB} \cite{shi2023mgtab}. For {Cresci-15} and {Twibot-20} we adopt the user-profile and textual features employed in ref. \cite{qiao2024dispelling}, whereas for {MGTAB} we follow the extraction protocol of ref. \cite{yang2024sebot}. Users are modeled as nodes and their follower/friend links as edges of the social graph. Following \cite{qiao2024dispelling}, each dataset is divided into 70\% training, 20\% validation, and 10\% test splits. We report \textit{Accuracy} and \textit{F1-score} as our primary metrics.

\textbf{Baselines.} We compare RABot against both feature-based and graph-based baselines. The feature-based counterparts include Wei \textit{et al.} \cite{wei2019twitter}, Varol \textit{et al.} \cite{varol2017online}, SATAR \cite{feng2021satar}, and SGBot \cite{yang2020scalable}. The graph-based competitors comprise GCN \cite{kipf2016semi}, {GAT} \cite{brody2021attentive}, {RGT} \cite{feng2022heterogeneity}, {BotRGCN} \cite{feng2021botrgcn}, {BECE} \cite{qiao2024dispelling}, {LMBot} \cite{cai2024lmbot}, {BotDGT} \cite{He2024Dynamicity}, and {SEBot} \cite{yang2024sebot}.

\textbf{Implementation Details.} We implement the proposed model RABot in PyTorch with Geometric. The GNN classifier can plug in any mainstream GNN backbone; unless stated otherwise, we employ the default architecture described in Section 4. Optimization is performed with Adam \cite{kingma2014adam} for 300 epochs and a learning rate of 0.001. All experiments run on a server with two NVIDIA RTX A800 GPUs (80\,GB each), a 16-core CPU, and 503\,GB RAM. Training completes in roughly 2, 5, and 10 minutes on {Cresci-15}, {Twibot-20}, and {MGTAB} datasets, respectively.

\begin{figure*}[!t]
    \centering
    \begin{subfigure}[t]{0.31\textwidth}
        \includegraphics[width=\textwidth]{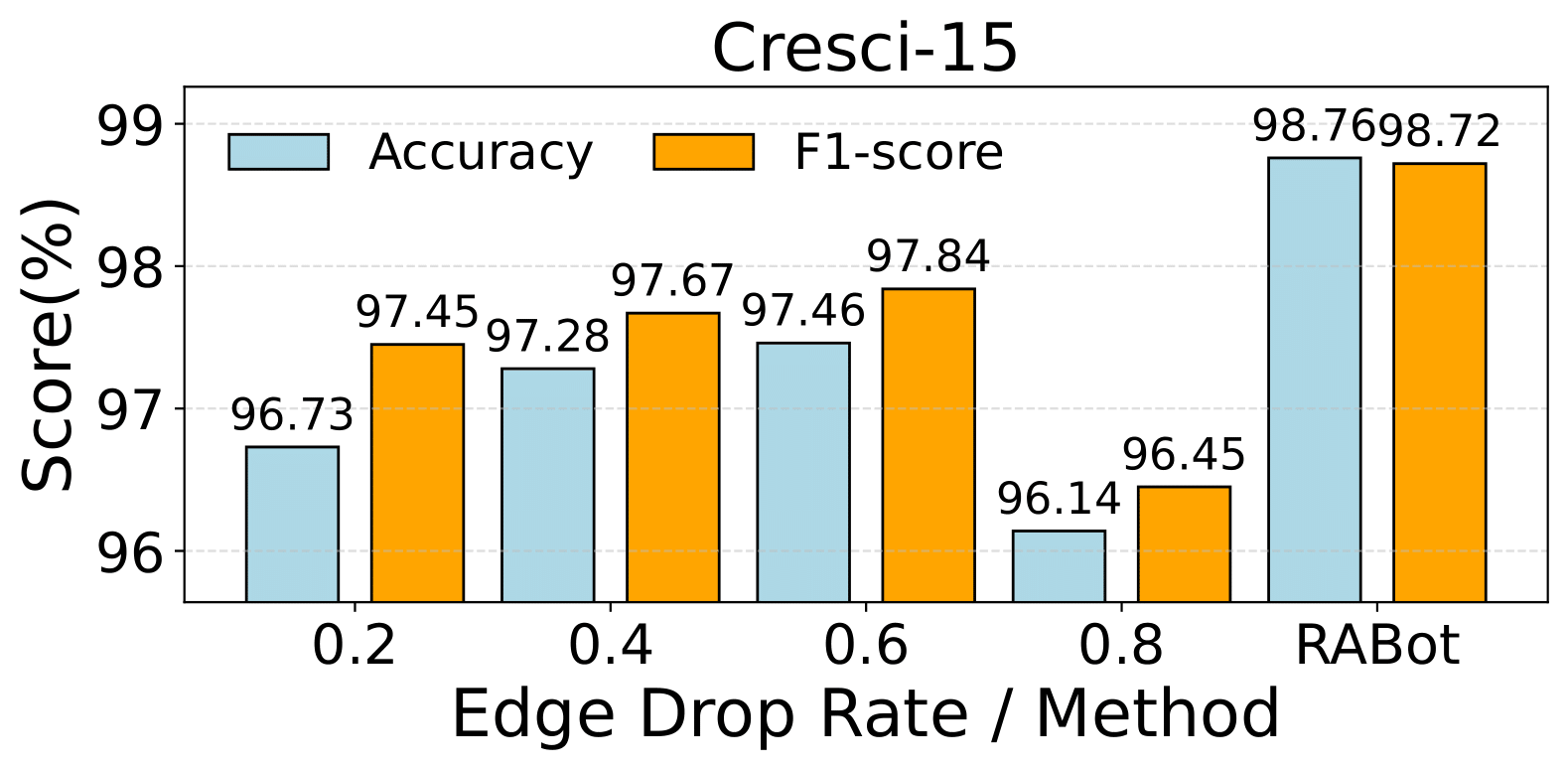}
    \end{subfigure}
    \begin{subfigure}[t]{0.31\textwidth}
        \includegraphics[width=\textwidth]{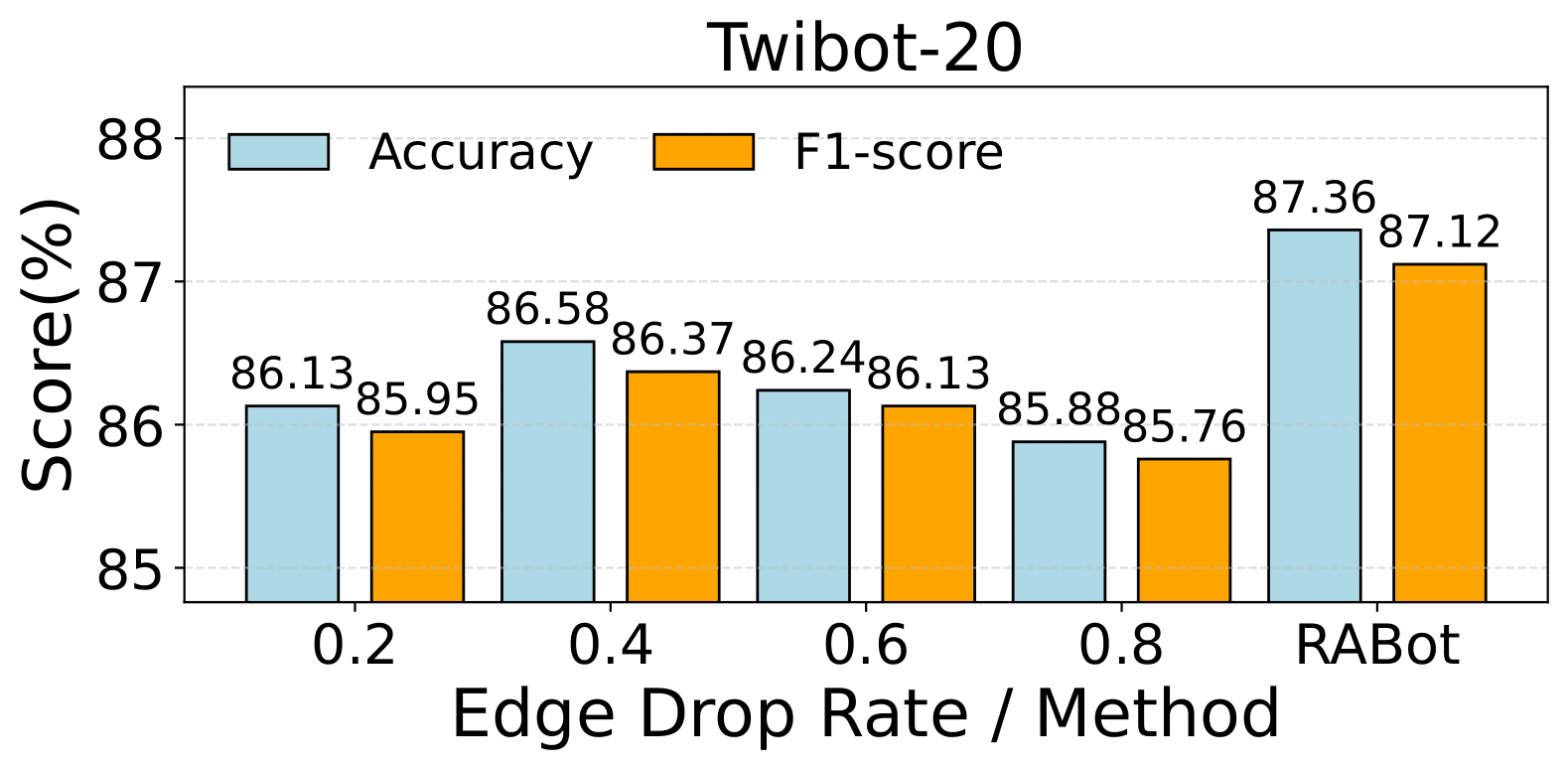}
    \end{subfigure}
    \begin{subfigure}[t]{0.31\textwidth}
        \includegraphics[width=\textwidth]{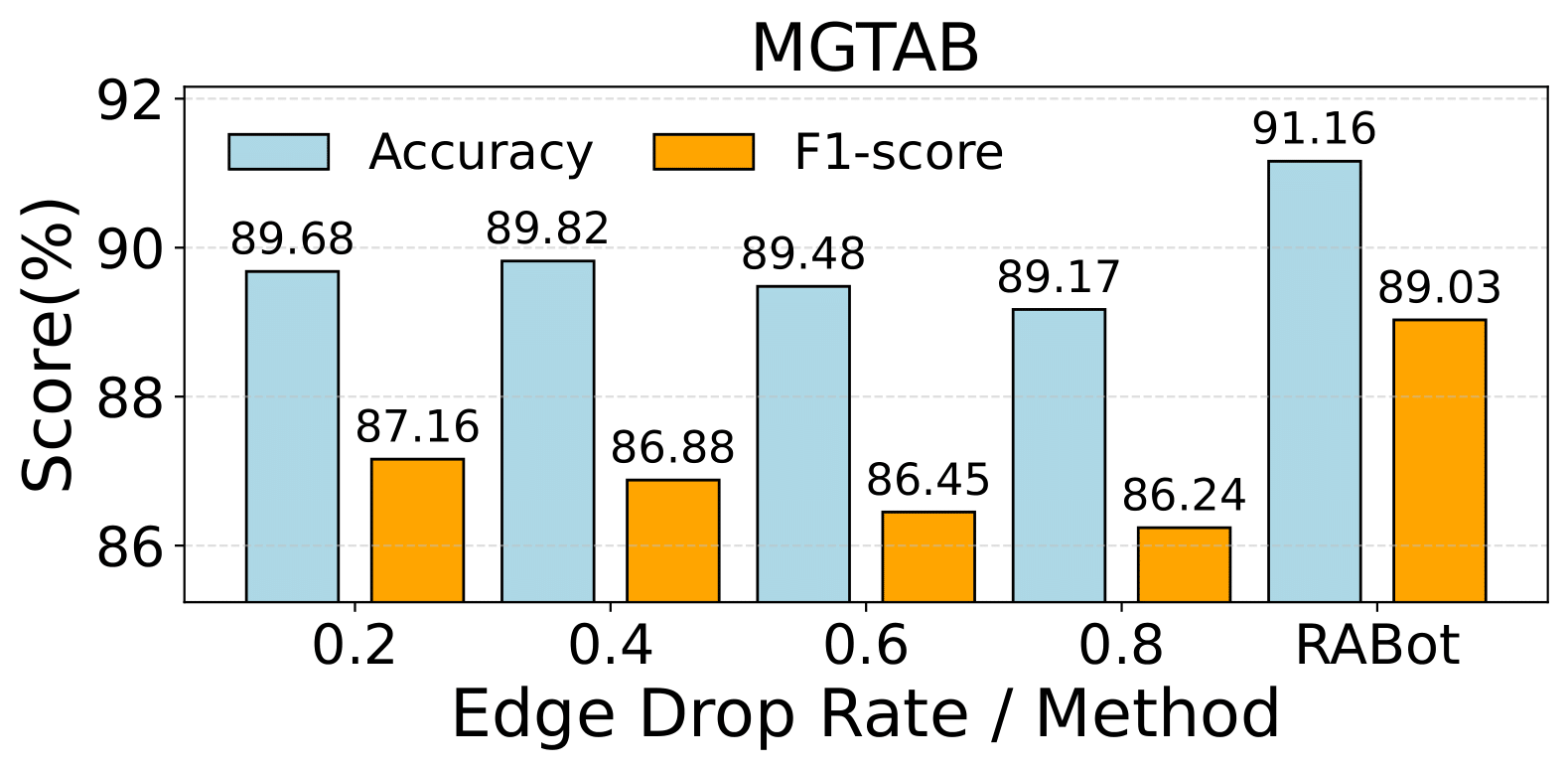}
    \end{subfigure}
 
    \caption{Performance of RABot with a RGCN backbone versus random edge removal of different drop rates on the Cresci-15, Twibot-20, and MGTAB datasets.}
    \label{fig1}
    \vspace{-5pt}
\end{figure*}

\subsection{Main Results}

Table~\ref{tab:modelCompare} presents the averages (\textit{$\pm$\,std}) of five independent runs of the proposed method and the compared baselines. Below we highlight four key observations.

\paragraph{Overall superiority.}
RABot establishes a new state of the art on \emph{all} three benchmarks.  
On Cresci-15, RABot\,(RGT) attains 99.14$\pm$0.21\% accuracy and 98.94$\pm$0.34\% F1-score, exceeding the runner-up BECE by 0.41 and 0.37 points, respectively.  
On Twibot-20, the same backbone yields 87.92$\pm$0.12\% accuracy and 88.40$\pm$0.28\% F1-score, surpassing the strongest baseline (SEBot) by 0.66 and 0.34 points.  
On the large-scale MGTAB corpus, RABot\,(RGCN) achieves 91.16$\pm$0.31\% accuracy and 89.03$\pm$0.28\% F1-score, yielding absolute gains of 0.85 and 0.93 points over the best prior model.

\paragraph{Backbone-agnostic gains.}
Irrespective of the underlying GNN, RABot delivers consistent improvements: on average across the three datasets it boosts GCN by +2.99/\,+2.64, GAT by +2.74/\,+2.59, RGT by +1.51/\,+1.42, and RGCN by +2.25/\,+2.00 percentage points in accuracy and F1-score, respectively. This backbone-agnostic behavior confirms that our class-balancing and RL-based edge-filtering modules are complementary to existing message-passing schemes.

\paragraph{Scalability to noisy, large graphs.}  
Performance gains are most pronounced on MGTAB, whose 1.7M edges exacerbate class imbalance and topological noise. Here RABot\,(GCN) registers the largest jump, indicating that our method is particularly beneficial when the base architecture is relatively weak and the graph is both large and noisy.

\paragraph{Improved stability.}  
Besides higher means, RABot reduces run-to-run variance. For instance, on MGTAB the standard deviation of F1-score of RGT drops from 1.12\% to 0.16\% after integrating RABot, suggesting a smoother optimization trajectory and better convergence.

In short, these findings verify that RABot not only pushes raw accuracy/F1-score forward but also provides a robust, plug-and-play enhancement for diverse GNN backbones under varied scenarios.

\begin{figure}[h]
    \centering
    \includegraphics[width=0.46\textwidth]{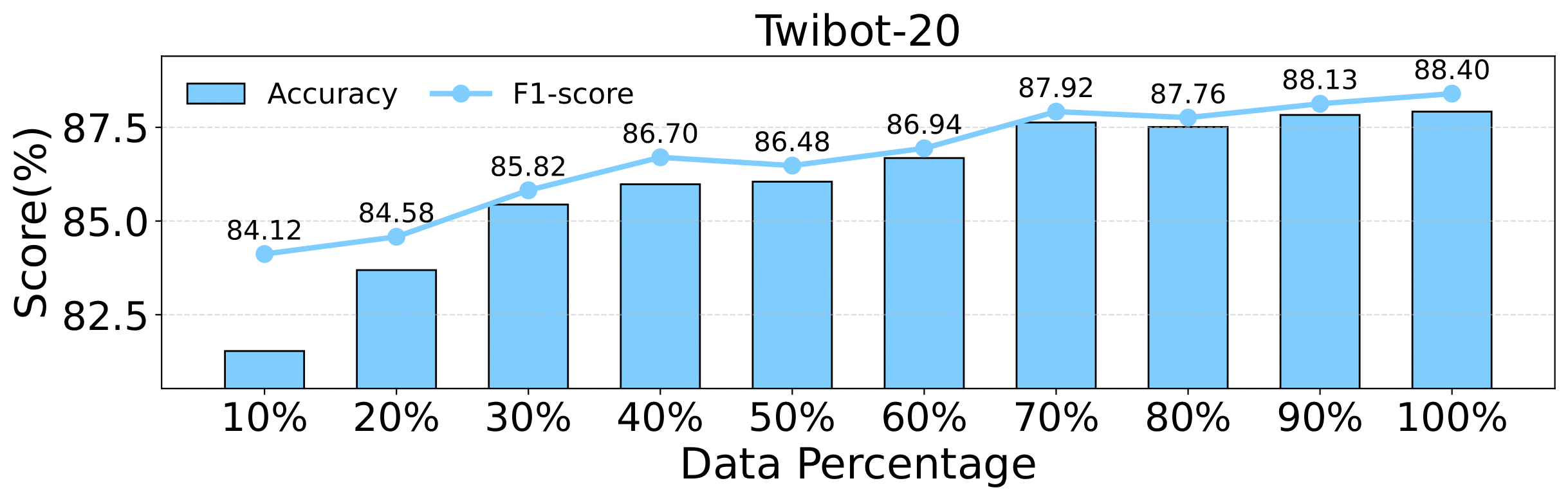}
    \hfill 
    \includegraphics[width=0.46\textwidth]{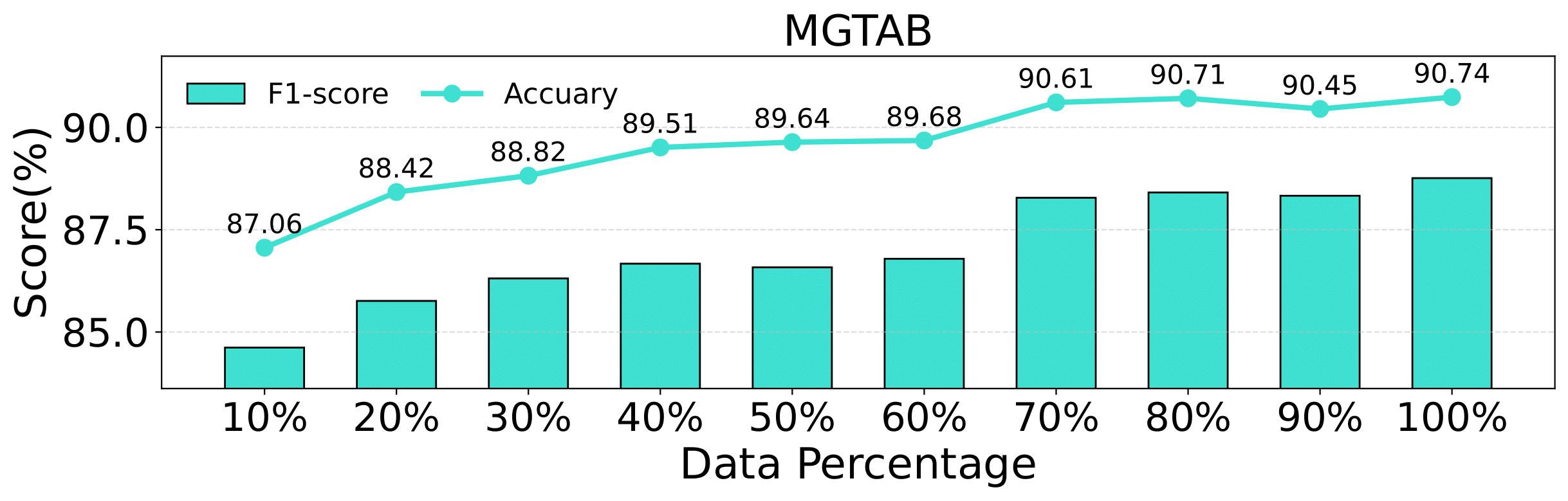}

    \caption{Data-efficiency study. The model is trained on random subsamples of the original training data ($10\%$-$100\%$) from Twibot-20 and MGTAB, and evaluates on the full test sets.}
    \label{fig_data_percentage}
    \vspace{-5pt}
\end{figure}

\begin{table}[h]
    \renewcommand{\arraystretch}{0.9}  
    \centering
     \small  
      \setlength{\tabcolsep}{3.5pt} 
    \begin{tabular}{@{}c|ll|ll@{}}
    \specialrule{1.2pt}{0pt}{0pt}  
      \multicolumn{1}{c|}{\multirow{2}{*}{\centering\raisebox{-1.2ex}{Ablation }}} &
      \multicolumn{2}{c|}{\centering \raisebox{-0.3ex}{Twibot-20}} &
      \multicolumn{2}{c}{\centering \raisebox{-0.3ex}{MGTAB}} \\ \cmidrule(l){2-5}
      &
      \multicolumn{1}{c}{Accuracy} &
      \multicolumn{1}{c|}{F1-score} &
      \multicolumn{1}{c}{Accuracy} &
      \multicolumn{1}{c}{F1-score} \\ \midrule

    \multicolumn{5}{c}{\textbf{RGT backbone}} \\ 
    \midrule
    RABot &
      \textbf{87.92$\pm$0.12} & \textbf{88.40$\pm$0.28} &
      \textbf{90.74$\pm$0.15} & \textbf{88.76$\pm$0.16} \\
    
    w/o MA&

      87.40$\pm$0.27 & 88.08$\pm$0.17 &
      90.42$\pm$0.31 & 88.41$\pm$0.29 \\

    w/o FA &

      87.31$\pm$0.16 & 87.96$\pm$0.31 &
      89.90$\pm$0.19 & 86.88$\pm$0.23 \\
    
    w/o EF &

      86.96$\pm$0.12 & 87.74$\pm$0.34 &
      89.63$\pm$0.18 & 86.96$\pm$0.37 \\
    
    w/o GC &

      85.18$\pm$0.19 & 86.83$\pm$0.22 &
      87.23$\pm$0.24 & 85.93$\pm$0.38 \\
    
      \midrule

    \multicolumn{5}{c}{\textbf{RGCN backbone}} \\ 
    \midrule
    
    RABot &
     
      \textbf{87.36$\pm$0.28} & \textbf{87.12$\pm$0.19} &
      \textbf{91.16$\pm$0.31} & \textbf{89.03$\pm$0.28} \\
    
    w/o MA &

      86.89$\pm$0.52 & 86.75$\pm$0.38 &
      90.91$\pm$0.43 & 88.90$\pm$0.44 \\
    
    w/o FA &
     
      86.81$\pm$0.47 & 86.62$\pm$0.26 &
      89.88$\pm$0.45 & 87.28$\pm$0.47 \\
    
    w/o EF &
     
      86.03$\pm$0.27 & 86.14$\pm$0.43 &
      89.73$\pm$0.39 & 87.31$\pm$0.40 \\
    
    w/o GC &
     
      85.46$\pm$0.24 & 85.68$\pm$0.16 &
      87.04$\pm$0.58 & 86.46$\pm$0.47 \\
    
    \specialrule{1.2pt}{0pt}{0pt}  
    \end{tabular}%
    \caption{Ablation study on the Twibot-20 and MGTAB datasets. ``w/o'' indicates that the specified component is removed. MA, FA, EF, and GC denote the multi-head attention, feature-augmentation, enhanced edge-filtering, and GNN classification modules, respectively.}
    \label{tab:modelCompare2}
\end{table}

\subsection{Ablation Study}
\label{sec:ablation}
To quantify the impact of each design choice, we systematically disable or replace individual components of RABot and re-evaluate the model on two relatively large-scale benchmarks.  
The ablated variants include (i) w/o~MA: removing the multi-head attention (MA) in the node encoder; (ii) w/o~FA: removing the feature-augmentation (FA) module;  
(iii) w/o~EF: omitting the enhanced edge-filtering (EF) mechanism; and  
(iv) w/o~GC: discarding the GNN classification (GC) module in the propagation stage.  
Results for both the RGT and RGCN backbones are summarized in Table~\ref{tab:modelCompare2}.

\noindent\textbf{Full model performs best.}  
Across all datasets and backbones, the complete RABot achieves the highest accuracy and F1-score, confirming the synergistic effect of its components.

\noindent\textbf{Multi-head attention matters.}  
Removing MA causes noticeable yet moderate drops (e.g., $-0.52\%$ of accuracy on Twibot-20 and $-0.32\%$ of F1-score on MGTAB with the RGT backbone), indicating that attention better captures heterogeneous user signals than a linear projection.

\noindent\textbf{Feature augmentation aids class balance.}  
Eliminating FA consistently harms performance, particularly on MGTAB where long-tailed class distribution is severe (accuracy decreases from 90.74\% to 89.90\% for RGT and from 91.16\% to 89.88\% for RGCN).  
This underscores the role of FA in mitigating imbalance and enriching minority representations.

\noindent\textbf{Edge filtering is critical.}  
Removing EF yields the largest single-component degradation aside from GC, with average accuracy/F1-score losses of $1.21$/$1.29$ points across datasets. The decline is most evident on MGTAB ($-1.43$ on accuracy, RGCN), demonstrating that denoising spurious edges is essential when social graphs contain deceptive connections.

\noindent\textbf{GNN classification drives the biggest gains.}  
Excluding GC results in the steepest drop, up to $-3.51\%$ regarding accuracy and $-2.83\%$ of F1-score on MGTAB (RGT), highlighting its importance for refining node features during message passing.

\subsection{Discussions}
\label{sec:vis}

\paragraph{Edge-filtering efficacy.}  
To validate the proposed edge-filtering strategy, we compare RABot with a baseline that randomly removes the same proportion of edges.  
Figure~\ref{fig1} shows that RABot consistently surpasses random deletion across all datasets.  
The gains stem from the similarity-aware edge features and the reinforcement-learning filter, which jointly suppress structural noise and strengthen the GNN's ability to capture salient connectivity patterns.

\begin{table}[H]
    \centering
    \small 
    \renewcommand{\arraystretch}{0.9}  
    \setlength{\tabcolsep}{6pt}

    \begin{tabular}{@{}c|cc|cc@{}}
    \specialrule{1.2pt}{0pt}{0pt}  
      \multicolumn{1}{c|}{\multirow{2}{*}{\centering\raisebox{-1.2ex}{Thresholds}}} &
      \multicolumn{2}{c|}{\centering \raisebox{-0.3ex}{Twibot-20}} &
      \multicolumn{2}{c}{\centering \raisebox{-0.3ex}{MGTAB}} \\ \cmidrule(l){2-5}
      &
      \multicolumn{1}{c}{Accuracy} &
      \multicolumn{1}{c|}{F1-score} &
      \multicolumn{1}{c}{Accuracy} &
      \multicolumn{1}{c}{F1-score} \\ \midrule
    
    0.2 &
      86.16 & 86.04 &
      89.97 & 87.33 \\
    
    0.4&
      86.62 & 86.48 &
      90.54 & 88.15 \\
    
    0.6 &
      86.73 & 86.61 &
      90.32 & 87.84 \\
    
    0.8 &
      86.06 & 85.94 &
      89.89 & 87.10 \\
    
    \textbf{ours} &
    \textbf{87.36} & \textbf{87.12} &
    \textbf{91.16} & \textbf{89.03} \\
    \midrule
    \specialrule{0.8pt}{0pt}{0pt}  
    \end{tabular}

    \caption{Performance comparison between fixed and dynamic filtering thresholds (ours) on the Twibot-20 and MGTAB datasets.}
    \label{tab:fixedanddynamic}
\end{table}


\paragraph{Performance under limited supervision.} 
Real-world social networks often have sparse and uneven annotations, challenging model generalization.  
To assess robustness in such settings, we vary the proportion of labelled training data from 10\,\% to 100\,\% (test sets remain fixed) and evaluate RABot\,(RGT) on Twibot-20 and MGTAB.  
Figure~\ref{fig_data_percentage} indicates that RABot retains strong performance even with substantially reduced supervision:  
on Twibot-20 it outperforms most baselines using only 80\,\% of the training data, and on MGTAB it surpasses the state-of-the-art RGT backbone with just 50\,\%.  
When the training ratio exceeds 70\,\%, gains saturate, implying that RABot already captures the critical patterns in low- and medium-resource regimes.

\begin{figure}[h]
    \centering
    \begin{subfigure}[t]{0.23\textwidth}
        \includegraphics[width=\textwidth]{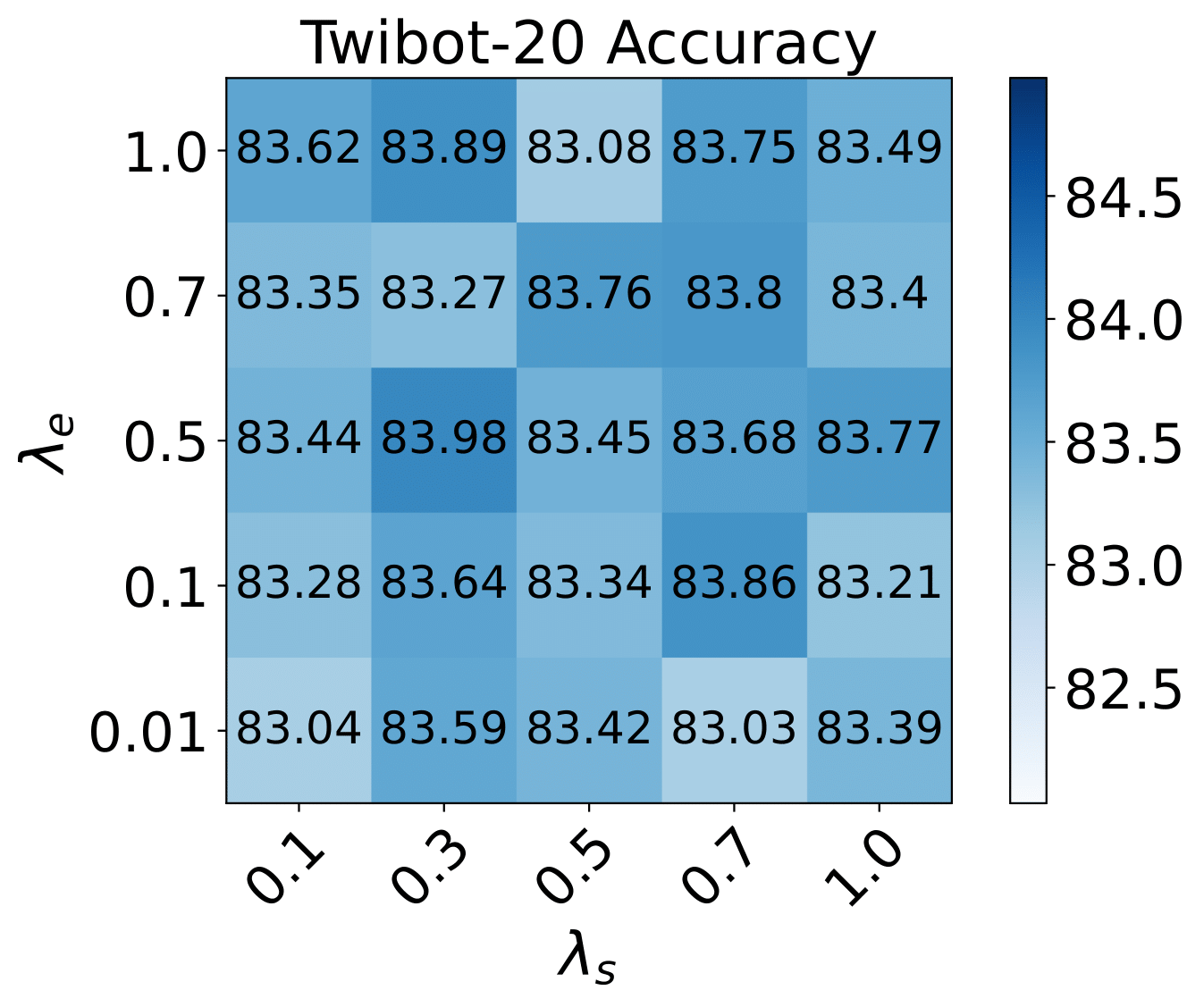}
    \end{subfigure}
    \begin{subfigure}[t]{0.23\textwidth}
        \includegraphics[width=\textwidth]{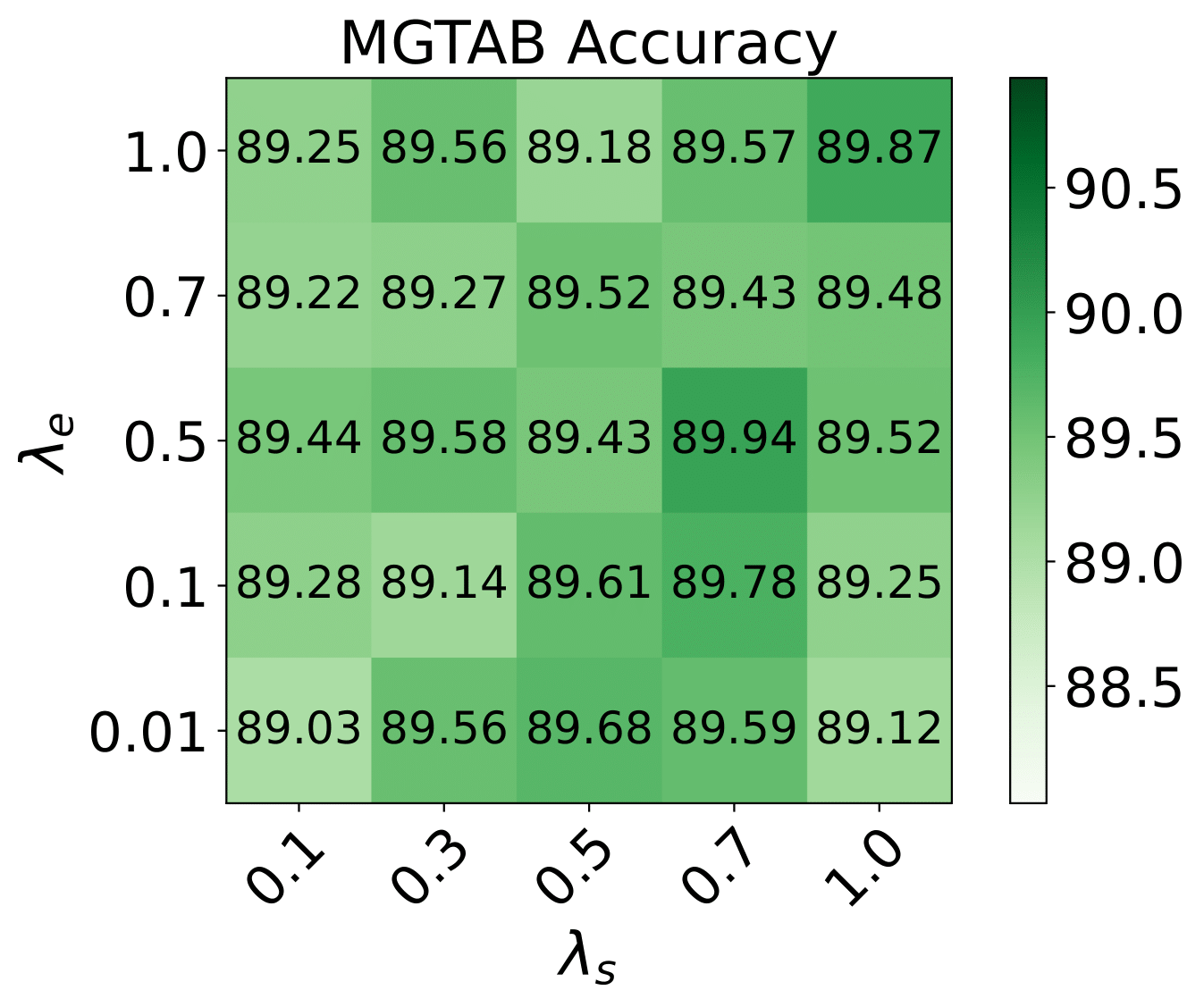}
    \end{subfigure}
  
    \begin{subfigure}[t]{0.23\textwidth}
        \includegraphics[width=\textwidth]{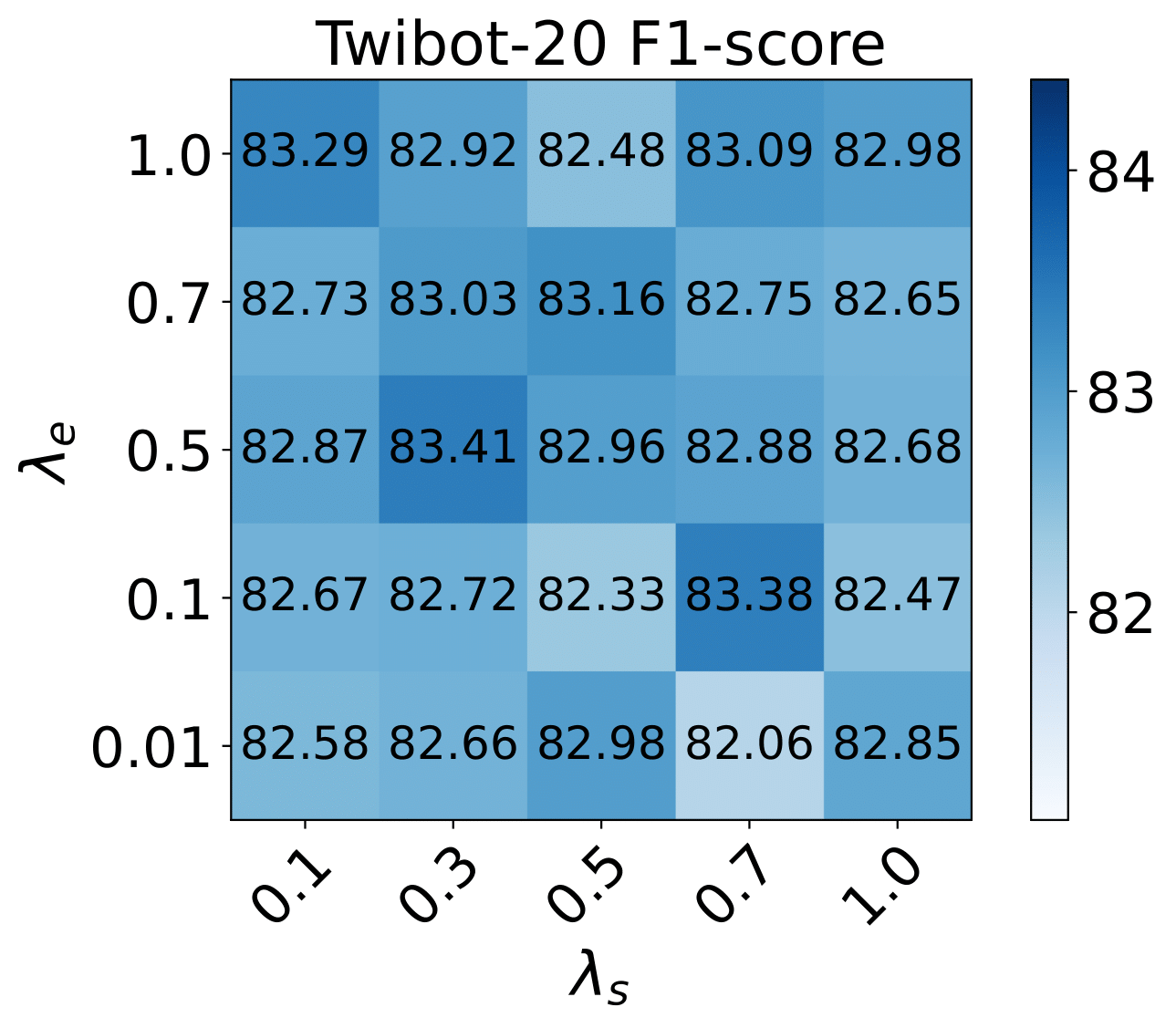}
    \end{subfigure}
    \begin{subfigure}[t]{0.23\textwidth}
        \includegraphics[width=\textwidth]{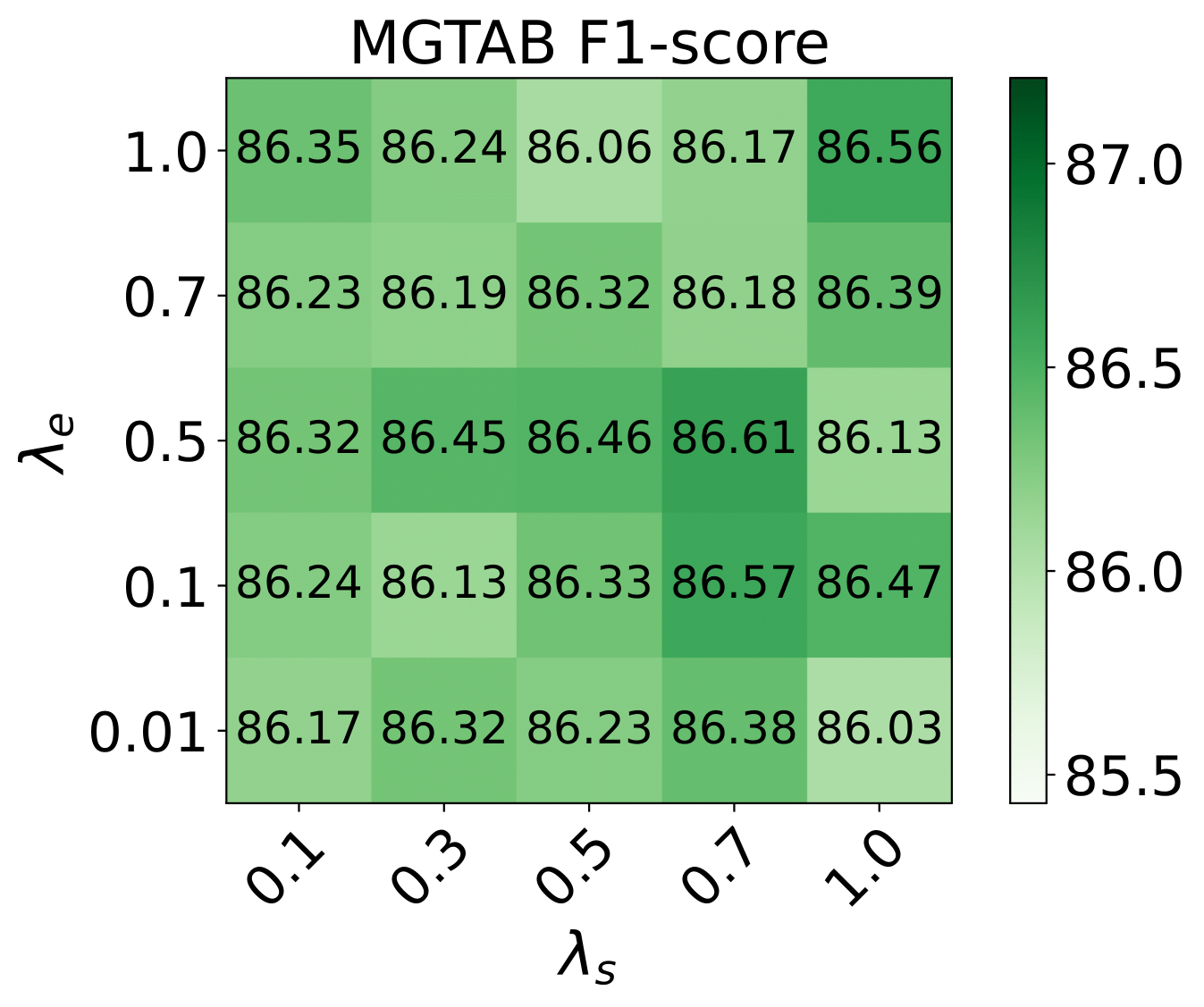}
    \end{subfigure}
    \caption{Sensitivity of RABot with a GAT backbone to the loss weights $\lambda_s$ and $\lambda_e$ on the Twibot-20 and MGTAB.}
    \label{fighotmap}
    \vspace{-5pt}
\end{figure}

\paragraph{Hyperparameter sensitivity.}  
We further show the impact of the loss weights $\lambda_s$ and $\lambda_e$ in Figure \ref{fighotmap}. Across the entire grid, accuracy variations are all less than $1\%$ on Twibot-20 and MGTAB. Such narrow ranges confirm that RABot remains stable even when the two coefficients deviate considerably from their optimal values. The best results on both datasets appear when $\lambda_e$ lies between 0.5 and 0.7. Smaller values (\,$\le$\,0.1) do not provide a strong enough signal for edge cleaning, whereas very large values (1.0) slightly over-emphasize the edge term and reduce node-level accuracy. 

In addition, a moderate emphasis on synthetic minority nodes is beneficial. For Twibot-20, the highest scores are reached at $\lambda_s=0.3$ and $\lambda_s=0.5$, respectively; for MGTAB the optimum shifts to $\lambda_s=0.7$, which reflects the heavier class imbalance in that dataset. The F1-score follows the same trend as accuracy, indicating a consistent trade-off between precision and recall. Besides, the proposed method is also robust to the number of GNN layers.

\paragraph{Effect of dynamic filtering threshold.} 
To assess the impact of dynamic thresholding, we compare our method with fixed thresholds on Twibot-20 and MGTAB.  
As shown in Table~\ref{tab:fixedanddynamic}, RABot (RGCN) yields the highest accuracy and F1-score on both datasets, outperforming all fixed settings.  
Compared to the best fixed threshold, it improves accuracy by 0.63\% on Twibot-20 and 0.62\% on MGTAB.

\begin{table}[h]
\centering
\small
\renewcommand{\arraystretch}{0.9} 
\setlength{\tabcolsep}{3pt}  
\begin{tabular}{lcccc}
\specialrule{0.4pt}{0pt}{0pt}
\toprule
Methods & Epoch Time & Epoch & Training Time & Peak Memory \\
\midrule
BECE & 8.4s & 100 & 14min & 8268M \\
LMBot & 4.2s & 300 &21min & 18636M \\
BotDGT & 27s & \textbf{20} &9min & 5724M \\
SEBot & 13.5s & 80 &18min & 14482M \\
\midrule
RABot & \textbf{1.0s} & 300 & \textbf{5min} & \textbf{2614M} \\
\bottomrule
\specialrule{0.4pt}{0pt}{0pt} 
\end{tabular}
\caption{Comparisons of computational cost on Twibot-20.}
\label{tab:Runningtime}
\end{table}

\paragraph{Computational efficiency.} 
Efficient training is essential for real-world deployment.  
As shown in Table~\ref{tab:Runningtime}, RABot (RGT) achieves the shortest training time and lowest peak memory usage among all methods.  
Despite running for 300 epochs, it is over 4 times faster than LMBot and uses only 14\% of its memory.  
These results highlight RABot's superior efficiency and scalability.

\section{Conclusion}

We presented the Reinforcement-guided graph Augmentation social Bot detector (RABot), a novel framework for social bot detection that couples an oversampling-based feature-augmentation strategy with a reinforcement-learning-driven edge-filtering mechanism. By jointly refining node representations and graph topology, RABot adapts to diverse graph structures and achieves superior robustness. Extensive evaluations on multiple benchmarks confirm that RABot consistently outperforms state-of-the-art alternatives. 
Future work will focus on further improving the efficiency of the edge-filtering module, lowering computational overhead, and extending RABot to domains beyond social media.

\section{Acknowledgments}

This work was supported in part by the National Natural Science Foundation of China under Grant 62572404, Grant 62203363, and Grant 62402458; in part by the Guangdong Basic and Applied Basic Research Foundation under Grant No. 2025A1515011565.

\bibliography{aaai2026}

\end{document}